\documentclass[journal]{IEEEtran}
\usepackage{url}
\usepackage[utf8]{inputenc}
\usepackage{xcolor}
\usepackage{amsmath}
\usepackage{amssymb}

\usepackage[acronyms,nonumberlist,nopostdot,nomain,nogroupskip]{glossaries}
\usepackage{tablefootnote}
\usepackage{booktabs}
\usepackage{tabularx}
\usepackage{tikz}
\usepackage{pgfplots}
\pgfplotsset{compat=newest} 
\pgfplotsset{plot coordinates/math parser=false} 
\newlength\fheight
\newlength\fwidth
\usetikzlibrary{plotmarks,patterns,decorations.pathreplacing,backgrounds,calc,arrows,arrows.meta,spy,matrix}
\usepgfplotslibrary{patchplots,groupplots}
\usepackage{tikzscale}
\usepackage{hyperref}

\usepackage{multirow}
\usepackage{tkz-kiviat}

\usepackage[font=scriptsize]{subcaption}
\usepackage[font=footnotesize]{caption}

\usepackage{mathtools}

\usepackage{dblfloatfix}    % To enable figures at the bottom of page
\usepackage{colortbl}

\usepackage{makecell}
\usepackage{diagbox}
\usepackage{tikz-qtree}
\usetikzlibrary{trees} % this is to allow the fork right path
\newacronym{3gpp}{3GPP}{3rd Generation Partnership Project}
\newacronym{adc}{ADC}{Analog to Digital Converter}
\newacronym{5g}{5G}{5th generation}
\newacronym{aimd}{AIMD}{Additive Increase Multiplicative Decrease}
\newacronym{am}{AM}{Acknowledged Mode}
\newacronym{amc}{AMC}{Adaptive Modulation and Coding}
\newacronym{aqm}{AQM}{Active Queue Management}
\newacronym{awgn}{AGWN}{Additive White Gaussian Noise}
\newacronym{balia}{BALIA}{Balanced Link Adaptation}
\newacronym{bdp}{BDP}{Bandwidth-Delay Product}
\newacronym{bf}{BF}{beamforming}
\newacronym{cc}{CC}{Congestion Control}
\newacronym{cdf}{CDF}{Cumulative Distribution Function}
\newacronym{cn}{CN}{Core Network}
\newacronym{cqi}{CQI}{Channel Quality Information}
\newacronym{cp}{CP}{Control Plane}
\newacronym{csirs}{CSI-RS}{Channel State Information - Reference Signal}
\newacronym{dc}{DC}{Dual Connectivity}
\newacronym{dce}{DCE}{Direct Code Execution}
\newacronym{dci}{DCI}{Downlink Control Information}
\newacronym{dl}{DL}{Downlink}
\newacronym{dmr}{DMR}{Deadline Miss Ratio}
\newacronym{dmrs}{DMRS}{DeModulation Reference Signal}
\newacronym{e2e}{E2E}{End-to-End}
\newacronym{si}{SI}{Study Item}
\newacronym{ecn}{ECN}{Explicit Congestion Notification}
\newacronym{edf}{EDF}{Earliest Deadline First}
\newacronym{enb}{eNB}{evolved Node Base}
\newacronym{epc}{EPC}{Evolved Packet Core}
\newacronym{es}{ES}{Edge Server}
\newacronym{fdma}{FDMA}{Frequency Division Multiple Access}
\newacronym{fdd}{FDD}{Frequency Division Duplexing}
\newacronym[firstplural=Radio Access Technologies (RATs)]{rat}{RAT}{Radio Access Technology}
\newacronym[firstplural=Radio Access Technology (RTs)]{rt}{RT}{Radio Technology}
\newacronym{fs}{FS}{Fast Switching}
\newacronym{ftp}{FTP}{File Transfer Protocol}
\newacronym{gnb}{gNB}{Next Generation Node Base}
\newacronym{harq}{HARQ}{Hybrid Automatic Repeat reQuest}
\newacronym{hetnet}{HetNet}{Heterogeneous Network}
\newacronym{hh}{HH}{Hard Handover}
\newacronym{hol}{HOL}{Head-of-Line}
\newacronym{ia}{IA}{Initial Access}
\newacronym{imt}{IMT}{International Mobile Telecommunication}
\newacronym{iot}{IoT}{Internet of Things}
\newacronym{los}{LOS}{Line of Sight}
\newacronym{lte}{LTE}{Long Term Evolution}
\newacronym{m2m}{M2M}{Machine to Machine}
\newacronym{mac}{MAC}{Medium Access Control}
\newacronym{mc}{MC}{Multi-Connectivity}
\newacronym{mcs}{MCS}{Modulation and Coding Scheme}
\newacronym{mec}{MEC}{Mobile Edge Cloud}
\newacronym{mi}{MI}{Mutual Information}
\newacronym{mimo}{MIMO}{Multiple Input, Multiple Output}
\newacronym{mmwave}{mmWave}{millimeter wave}
\newacronym{mptcp}{MPTCP}{Multipath TCP}
\newacronym{mr}{MR}{Maximum Rate}
\newacronym{mss}{MSS}{Maximum Segment Size}
\newacronym{mtd}{MTD}{Machine-Type Device}
\newacronym{mtu}{MTU}{Maximum Transmission Unit}
\newacronym{nfv}{NFV}{Network Function Virtualization}
\newacronym{nlos}{NLOS}{Non Line of Sight}
\newacronym{nlosb}{NLOSb}{Building Non Line of Sight}
\newacronym{nlosv}{NLOSv}{Vehicle Non Line of Sight}
\newacronym{nr}{NR}{New Radio}
\newacronym{ofdm}{OFDM}{Orthogonal Frequency Division Multiplexing}
\newacronym{pdcch}{PDCCH}{Physical Downlonk Control Channel}
\newacronym{pdcp}{PDCP}{Packet Data Convergence Protocol}
\newacronym{pdsch}{PDSCH}{Physical Downlink Shared Channel}
\newacronym{pdu}{PDU}{Packet Data Unit}
\newacronym{pf}{PF}{Proportional Fair}
\newacronym{pgw}{PGW}{Packet Gateway}
\newacronym{phy}{PHY}{Physical}
\newacronym{pbch}{PBCH}{Physical Broadcast Channel}
\newacronym[plural=\gls{mme}s,firstplural=Mobility Management Entities (MMEs)]{mme}{MME}{Mobility Management Entity}
\newacronym{prb}{PRB}{Physical Resource Block}
\newacronym{pss}{PSS}{Primary Synchronization Signal}
\newacronym{pucch}{PUCCH}{Physical Uplink Control Channel}
\newacronym{pusch}{PUSCH}{Physical Uplink Shared Channel}
\newacronym{rach}{RACH}{Random Access Channel}
\newacronym{ran}{RAN}{Radio Access Network}
\newacronym{red}{RED}{Random Early Detection}
\newacronym{rf}{RF}{Radio Frequency}
\newacronym{rlc}{RLC}{Radio Link Control}
\newacronym{rlf}{RLF}{Radio Link Failure}
\newacronym{rrc}{RRC}{Radio Resource Control}
\newacronym{rrm}{RRM}{Radio Resource Management}
\newacronym{rr}{RR}{Round Robin}
\newacronym{rs}{RS}{Remote Server}
\newacronym{rsrp}{RSRP}{Reference Signal Received Power}
\newacronym{rss}{RSS}{Received Signal Strength}
\newacronym{rtt}{RTT}{Round Trip Time}
\newacronym{rw}{RW}{Receive Window}
\newacronym{rx}{RX}{Receiver}
\newacronym{sa}{SA}{standalone}
\newacronym{sack}{SACK}{Selective Acknowledgment}
\newacronym{sap}{SAP}{Service Access Point}
\newacronym{sch}{SCH}{Secondary Cell Handover}
\newacronym{scoot}{SCOOT}{Split Cycle Offset Optimization Technique}
\newacronym{sdma}{SDMA}{Spatial Division Multiple Access}
\newacronym{sinr}{SINR}{Signal to Interference plus Noise Ratio}
\newacronym{sm}{SM}{Saturation Mode}
\newacronym{snr}{SNR}{Signal to Noise Ratio}
\newacronym{son}{SON}{Self-Organizing Network}
\newacronym{ss}{SS}{Synchronization Signal}
\newacronym{srs}{SRS}{Sounding Reference Signal}
\newacronym{sss}{SSS}{Secondary Synchronization Signal}
\newacronym{tb}{TB}{Transport Block}
\newacronym{tcp}{TCP}{Transmission Control Protocol}
\newacronym{tdd}{TDD}{Time Division Duplexing}
\newacronym{tdma}{TDMA}{Time Division Multiple Access}
\newacronym{tfl}{TfL}{Transport for London}
\newacronym{tm}{TM}{Transparent Mode}
\newacronym{prr}{PRR}{Packet Reception Ratio}
\newacronym{trp}{TRP}{Transmitter Receiver Pair}
\newacronym{tti}{TTI}{Transmission Time Interval}
\newacronym{ttt}{TTT}{Time-to-Trigger}
\newacronym{tx}{TX}{Transmitter}
\newacronym{ue}{UE}{User Equipment}
\newacronym{ul}{UL}{Uplink}
\newacronym{uml}{UML}{Unified Modeling Language}
\newacronym{um}{UM}{Unacknowledged Mode}
\newacronym{utc}{UTC}{Urban Traffic Control}
\newacronym{vm}{VM}{Virtual Machine}
\newacronym{rsrq}{RSRQ}{Reference Signal Received Quality}
\newacronym{rssi}{RSSI}{Received Signal Strength Indicator}
\newacronym{crs}{CRS}{Cell Reference Signal}
\newacronym{v2v}{V2V}{Vehicle-to-Vehicle}
\newacronym{v2i}{V2I}{Vehicle-to-Infrastructure}
\newacronym{v2n}{V2N}{Vehicle-to-Network}
\newacronym{v2x}{V2X}{Vehicle-to-Everything}
\newacronym{vn}{VN}{Vehicular Node}
\newacronym{dsrc}{DSRC}{Dedicated Short Range Communication}
\newacronym{ci}{CI}{context information}
\newacronym{voi}{VoI}{value of information}
\newacronym{gps}{GPS}{Global Positioning System}
\newacronym{qos}{QoS}{Quality of Service}
\newacronym{ml}{ML}{Machine Learning}
\newacronym{ahp}{AHP}{Analytic Hierarchy Process}
\newacronym{lidar}{LIDAR}{Light Detection and Ranging}
\newacronym{sumo}{SUMO}{Simulation of Urban MObility}
\usepackage{cite}
\usepackage{bbold}
\usepackage{bbm}

\usepackage{array}
\newcolumntype{?}{!{\vrule width 1.5pt}}
\newcolumntype{P}[1]{>{\centering\arraybackslash}p{#1}}
\usepackage{makecell}
\usepackage{mdframed}
\usepackage[many]{tcolorbox}

\newtcbox{\mybox}[1][]{nobeforeafter,math upper,tcbox raise base,
  enhanced,frame hidden,boxrule=0pt,interior style={top color=green!10!white,
  bottom color=green!10!white,middle color=green!50!yellow},
  fuzzy halo=1pt with green,drop large lifted shadow,#1}

\usepackage{siunitx}
\usetikzlibrary{fadings}

\tikzfading[name=middle,
            top color=transparent!100,
            bottom color=transparent!100,
            middle color=transparent!20]

\usetikzlibrary{arrows,automata,calc,shapes, positioning,shadows,shadows.blur,shapes.geometric}

 \def \PLOS{
       \begin{equation}
       \begin{cases}
       	\text{PL}_{\rm LOS}^{\rm u}(d) = 38.77+16.7\log_{10}(d)+18.2\log_{10}(f_c)+\chi_a\\
       	\text{PL}_{\rm LOS}^{\rm h}(d) = 32.4+20\log_{10}(d)+20\log_{10}(f_c)+\chi_a
       	\label{eq:PL_LOS}
       	\end{cases}
       \end{equation} 
            }

\def \PNLOS{
                    \begin{equation}
       	\text{PL}_{\rm NLOS}(d) = 36.85 + 30\log_{10}(d) + 18.9\log_{10}(f_c)+\chi_a,
       \end{equation} 
       }

\makeglossaries

\begin{document}
\pagenumbering{gobble}

% \title{\huge A Comparison \hspace{-0.3cm} of IEEE~802.11p \hspace{-0.3cm} and \hspace{-0.3cm} mmWave Communications  to Support Future Vehicle-to-Vehicle Networking}

\title{\fontsize{23}{25}\selectfont Path Loss Models for  V2V mmWave Communication: Performance~Evaluation and Open Challenges}

\author{\vspace{-0.33cm}\IEEEauthorblockN{ Marco Giordani$^{\circ }$, Takayuki Shimizu$^{\dagger }$, Andrea Zanella$^{\circ }$, Takamasa Higuchi$^{\dagger }$, Onur Altintas$^{\dagger}$, Michele Zorzi$^{\circ }$}
\IEEEauthorblockA{\\
% $^{\circ }$Consorzio Futuro in Ricerca (CFR) and  University of Padova, Italy \\
%  email:\texttt{\{giordani,zanella,zorzi\}@dei.unipd.it, }\\
% $^{\dagger}$TOYOTA InfoTechnology Center,  U.S.A., Inc., Mountain View, CA\\
% email:\texttt{\{ta-higuchi,onur\}@us.toyota-itc.com }}}
$^{\circ }$Department of Information Engineering (DEI), University of Padova, Italy \\
$^{\dagger}$InfoTech Labs, Toyota Motor North America, Inc.\\
Email: {\{giordani,zanella,zorzi\}@dei.unipd.it}, {\{takayuki.shimizu,takamasa.higuchi,onur.altintas\}@toyota.com }\vspace{-0.99cm}}}

\maketitle

\begin{abstract}
Recently, millimeter wave (mmWave) bands have been investigated as a means to enhance automated driving and address the challenging data rate and latency demands of emerging automotive applications.
For the development of those systems to operate
in bands above 6 GHz, there is a need to have accurate channel models able to predict the peculiarities of the vehicular propagation at these bands, especially as far as Vehicle-to-Vehicle (V2V) communications are concerned.
In this paper, we validate the  channel model that the 3GPP has  proposed for NR-V2X systems, which (i) supports deployment scenarios for urban/highway propagation, and (ii) incorporates the effects of path loss, shadowing, line of sight probability, and static/dynamic blockage attenuation.
We also exemplify the impact of several automotive-specific parameters on the overall network performance considering realistic system-level simulation assumptions for typical scenarios.
Finally, we highlight potential inconsistencies of the model and provide recommendations for future measurement campaigns in vehicular~environments. 

%In this context, one of the most important points is to
\end{abstract}


\section{Introduction} % (fold)
\label{sec:introduction}

Over the past few years, the advance towards enhanced automated driving relying on high levels of connectivity has motivated the automotive industry to investigate
new radio access technologies as a tool to enable vehicular communications.
In particular, the \gls{mmwave} frequencies offer the potential of multi-gigabit-per-second data transmission rates and therefore represent a promising opportunity to support advanced automotive~applications~\cite{Boccardi,Heath_surveyV2X}.

In this regard, the \gls{3gpp} has recently developed some functionalities to provide enhancements of the traditional communication standards for both \gls{v2v} and \gls{v2n} communications, starting with the \gls{3gpp} Release-16 \gls{si} on NR-V2X~\cite{3GPP_RP181480}.
These developments include NR-based sidelink design and enhancements of Release-15 NR uplink/downlink for advanced V2X services in both sub-6 GHz and mmWave bands.
From a physical-layer perspective, the channel has been designed to be optimized for uses under 52.6 GHz and with the potential to be used for above 52.6 GHz (including therefore \gls{mmwave} transmissions)~\cite{3GPP_181435}.
However, frequencies above 52.6 GHz face more difficult challenges than conventional bands, e.g., severe propagation and penetration loss and strong power spectral density regulatory requirements~\cite{rappaport2014millimeter},  thereby calling for innovative channel modeling and design solutions~\cite{MOCAST_2017}.
While measurement campaigns in mmWave \gls{v2n} scenarios  have been quite widespread (e.g.,~\cite{kato2001its}),  measurements in the \gls{v2v} context are still very limited, and realistic scenarios are indeed hard to simulate. 
For example, Takahashi et al., in~\cite{takahashi2003distance},  have conducted measurements in crowded roads to characterize the mmWave propagation in \gls{v2v} networks, 
although validation through more accurate experimental equipment is needed.
Yamamoto et al., in~\cite{yamamoto2008path}, have also conducted some measurements to obtain a general  characterization of the V2V channel  at 60 GHz, but the model does not consider urban propagation, nor does it distinguish between static and dynamic blockage.
%In~\cite{schneider2000impact}, Schneider et al. have captured the effect of the road ondulation and reflectivity when considering V2V transmissions in the 77–81 GHz band, without deriving, however, a suitable path loss model.
The measurements in~\cite{38901} (with modified antenna heights) can be a starting point of discussion for the definition of a V2V channel model in above-6 GHz scenarios, although the parameters  are derived from cellular measurements which might not be fully representative of a vehicular system due to the more challenging propagation characteristics of highly mobile vehicular nodes. 
Some other measurements were conducted for frequencies between 450 MHz and 6 GHz, but did not include any empirical confirmation for use  at mmWaves~\cite{molisch2009survey}.

Along these lines, the 3GPP has recently specified how to model the \gls{v2v}  channel at mmWave frequencies~\cite{37885}.
In particular, distinctions between environmental and vehicular blockages,  as well as between urban and highway propagation scenarios, have been proposed.
However, before the model can be adopted as a feasible solution to simulate  V2V propagation, it is fundamental to evaluate its performance in realistic vehicular scenarios, a research issue that, to date, has not yet been thoroughly addressed.
In this paper,  we  provide the first numerical validation of the \gls{3gpp} \gls{v2v} path loss model for nodes operating at \glspl{mmwave}.
Moreover, we exemplify the impact of  several automotive-specific parameters, such as the antenna array size and the vehicular traffic density,  on the overall network performance, which is assessed in terms of packet reception ratio. 
To this end, we consider an extension of the 3GPP model, proposed in \cite{boban2016modeling}, which, unlike the 3GPP model, characterizes density-dependent propagation.
Finally, we highlight potential model limitations that should be overcome by future measurements in vehicular environments, such as the dynamics of the channel, the Doppler spread with directional antennas, the impact of the antenna placement, the traffic density, and the delay spread.

\renewcommand{\arraystretch}{0.7}
\begin{table*}[h!]
\centering
\footnotesize
 \caption{LOS and NLOSv  probabilities in  highway and urban scenarios according to the 3GPP model~\cite[Table 6.2-1]{37885}. The model does not distinguish between different densities of vehicular traffic nor does it define a closed-form expression for the NLOS probability$^{(*)}$.}
\label{tab:PL_state_3gpp}
\begin{tabular}{l?c?c?c}
\Xhline{2\arrayrulewidth}
\rule{0pt}{0.3cm} Path Loss Probability & \multicolumn{2}{c?}{Highway scenario} & Urban scenario  \rule[-0.2cm]{0pt}{0pt}                                                     \\ \Xhline{2\arrayrulewidth}
\multirow{2}{*}{$P_{\rm LOS}(d)$ } & $d\leq 475$ m              & $d> 475$ m     & \multirow{2}{*}{$\min\{1,1.05 e^{-0.0114d}\}$} \\ \cline{2-3}
& \begin{tabular}[c]{@{}c@{}}$\min\{1,(2.1013\cdot10^{-6})d^2-0.002d+1.0193\}$ \end{tabular} & \begin{tabular}[c]{@{}c@{}}$\max\{0,0.54-0.001(d-475)\}$\end{tabular}  &  \\ \hline
\rule{0pt}{0.3cm} $P_{\rm NLOSv}(d)$ &\multicolumn{3}{c}{$1-P_{\rm LOS}(d)$}\rule[-0.2cm]{0pt}{0pt}\\\Xhline{2\arrayrulewidth}
\multicolumn{4}{c}{\scriptsize\begin{tabular}[c]{@{}c@{}}$^{(*)}$The NLOS status is derived from  geometric considerations, which evaluate whether the direct path between  the TX and the RX is blocked by static obstructions, e.g., buildings.\end{tabular}} 
\end{tabular}
\vspace{-0.33cm}
\end{table*}

\renewcommand{\arraystretch}{0.7}
\begin{table*}[b!]
\centering
\footnotesize
 \caption{LOS, NLOSv and NLOS probabilities in  highway and urban scenarios, for different  traffic densities, according to the extended model in~\cite{boban2016modeling}$^{(*)}$.}
\label{tab:PL_state_boban}
\begin{tabular}{l?c?P{2.2cm}P{1.9cm}P{1.9cm}?c}
\Xhline{2\arrayrulewidth}
\rule{0pt}{0.2cm} \multirow{2}{*}{Probability} & \multirow{2}{*}{Density} & \multicolumn{3}{c?}{Highway scenario: $P_{*}(d) =\min\{1,\max\{0,ad^2+bd+c\}\}$}  &\multirow{2}{*}{Urban scenario: $P_{*}(d) = \min\{1,\max\{0,f(d)\}\}$}      \rule[-0.1cm]{0pt}{0pt}        \\ \cline{3-5}
& & a & b& c & \\ \Xhline{2\arrayrulewidth}
\rule{0pt}{0.2cm}  \multirow{3}{*}{$P_{\rm LOS}(d)$} & Low & $1.5\cdot10^{-6}$  & $-0.0015$  & $1$ & $0.8548\cdot e^{-0.0064 d}$  \\
& Medium & $2.7\cdot10^{-6}$  & $-0.0025$  & $1$ &  $0.8372\cdot e^{-0.0114 d}$ \\
& High & $3.2\cdot10^{-6}$  & $-0.003$  & $1$ & $0.8962\cdot e^{-0.0170 d}$ \rule[-0.1cm]{0pt}{0pt}  \\ \Xhline{2\arrayrulewidth}
  \rule{0pt}{0.2cm} \multirow{3}{*}{$P_{\rm NLOS}(d)$} & Low & $-2.9\cdot10^{-7}$ & $0.00059$ & $0.0017$ & \multirow{3}{*}{$1-P^{\rm u}_{\rm LOS}(d)-P^{\rm u}_{\rm NLOSv}(d)$} \\
 & Medium & $-3.7\cdot10^{-7}$ & $0.00061$ & $0.0150$ &  \\
 & High &  $-4.1\cdot10^{-7}$ & $0.00067$ & $0$ &  \rule[-0.1cm]{0pt}{0pt}  \\ \Xhline{2\arrayrulewidth}
 \rule{0pt}{0.2cm} \multirow{3}{*}{$P_{\rm NLOSv}(d)$} & Low &  \multicolumn{3}{c?}{\multirow{3}{*}{$1-P^{\rm h}_{\rm LOS}(d)-P^{\rm h}_{\rm NLOS}(d)$}} &  \hspace{0.37cm} $1/0.0396d \cdot \exp\left[-(\ln{(d)}-5.2718)^2/3.4827\right]$ \hspace{0.37cm} \\
 & Medium & & & & \hspace{0.37cm} $1/0.0312d\cdot \exp\left[-(\ln{(d)}-5.0063)^2/2.4544\right]$  \hspace{0.37cm} \\
  & High & & & & \hspace{0.37cm}  $1/0.0242d\cdot \exp\left[-(\ln{(d)}-5.0115)^2/2.2092\right]$ \hspace{0.37cm} \rule[-0.1cm]{0pt}{0pt} \\ 
  \Xhline{2\arrayrulewidth}
\multicolumn{6}{c}{\scriptsize\begin{tabular}[c]{@{}c@{}}$^{(*)}$The model in~\cite{boban2016modeling} extends the 3GPP model in~\cite{37885} to characterize the LOS, NLOSv and NLOS probabilities  as a function of the density of vehicular traffic.\end{tabular}} 

\end{tabular}
\end{table*}

\section{V2V mmWave Path Loss Modeling} % (fold)
\label{sec:3gpp_path_loss_model}

The first step towards proper vehicular protocol design is a deep understanding of the propagation model. 
In this section, we  describe the path loss characterization that the 3GPP is considering for  V2V \gls{mmwave}  communications. 

\subsection{LOS/NLOS Probability} % (fold)
\label{sub:los_nlos_probability}

In \gls{v2v} systems, the  path loss is modeled according to the following three  states~\cite{37885}:
\begin{enumerate}
	\item \emph{\gls{los}}, i.e., the propagation path is not blocked by vehicles nor environmental objects.
		\item \emph{\gls{nlosv}}, i.e.,  the \gls{los} path is  blocked by dynamic blockages (e.g., other vehicles).
	\item \emph{Non Line of Sight (NLOS)}, i.e., the \gls{los} path is  blocked by environmental blockages (e.g., buildings).
\end{enumerate}

The  \gls{los} and \gls{nlosv}  probabilities, i.e., $P_{\rm LOS}$ and  $P_{\rm NLOSv}$, are defined in~\cite[Table 6.2-1]{37885} and reported in Table~\ref{tab:PL_state_3gpp}. 

Although the model distinguishes between urban and highway scenarios (respectively denoted with superscripts $u$ and $h$ throughout the paper), it does not differentiate between different densities of vehicular traffic.
Moreover, the determination of the NLOS state is deterministic, i.e., it is based on purely geometric considerations which evaluate whether the V2V link is blocked or not by buildings, therefore a closed-form expression for the NLOS probability, i.e., $P_{\rm NLOS}$, is not currently provided.
As we will numerically demonstrate in Sec.~\ref{sub:performance_results}, such assumptions reduce the accuracy of the analysis and might result in misleading conclusions.
%Finally, the 3GPP does not distinguish between different densities of vehicular traffic and, as we will show in Sec.~\ref{sub:performance_results}, such assumptions significantly undermine the accuracy of the analysis and might lead to wrong conclusions.

\renewcommand{\arraystretch}{0.7}
\begin{table*}[t!]
\footnotesize
\centering
\caption{LOS, NLOSv and NLOS path loss equations for V2V links.}
\label{tab:PL}
\begin{tabular}{l?c?c?c}
\Xhline{2\arrayrulewidth}
\rule{0pt}{0.3cm}
Scenario & LOS path loss ($\text{PL}_{\rm LOS}$) & NLOSv path loss ($\text{PL}_{\rm NLOSv}$) & NLOS path loss ($\text{PL}_{\rm NLOS}$) \rule[-0.1cm]{0pt}{0pt}  \\ \hline
\Xhline{2\arrayrulewidth}
\rule{0pt}{0.3cm}
  Urban & $38.77+16.7\log_{10}(d)+18.2\log_{10}(f_c)+\chi_a$  & \multirow{2}{*}{PL$_{\rm LOS}+\mathcal{A}_{\rm NLOSv}$} & \multirow{2}{*}{$36.85 + 30\log_{10}(d) + 18.9\log_{10}(f_c) +\chi_a$}  \\
  Highway & $32.4+20\log_{10}(d)+20\log_{10}(f_c)+\chi_a$ & & \rule[-0.1cm]{0pt}{0pt} \\\hline
\Xhline{2\arrayrulewidth}        
\end{tabular}
\end{table*}

In this paper, we therefore consider an extension of the path loss probability equations presented above based on~\cite{boban2016modeling},  as summarized in Table~\ref{tab:PL_state_boban}. The model in~\cite{boban2016modeling}  (i) characterizes low, medium, and high densities of the vehicular traffic in both urban and highway scenarios, and (ii) introduces a probabilistic model for the NLOS probability  as a function of the inter-vehicle distance  (the longer the link, the more likely to intersect  one or more blockages).
In order to have realistic mobility traces for the vehicles in the considered environments, the authors in~\cite{boban2016modeling} have used \gls{sumo}~\cite{SUMO2012}, an open-source road traffic simulator designed to handle and model the traffic of large road networks.
The LOS/NLOSv/NLOS classification is finally provided by GEMV2, a freely available vehicular propagation modeling software which  performs geometry-based  blockage analyses based on the outlines of  buildings and vehicles. 
% In particular, when the LOS is obstructed by both environmental and dynamic objects, GEMV2 classifies this as NLOS state because environmental objects usually are the dominant  blocking factor.
% Moreover, since the presence of intermediate vehicles potentially obstructing the line of sight between the endpoints does not necessarily imply the \gls{nlosv} condition~\cite{boban2011impact}, GEMV2 calculates whether 60\% of the radius of the first Fresnel ellipsoid is free of  obstructions to decide  whether the LOS path is~blocked.

%Moreover, from an electromagnetic  perspective, the presence of intermediate vehicles potentially obstructing the line of sight between the endpoints does not necessarily imply the \gls{nlosv} condition. 
%It is also essential that the \emph{Fresnel ellipsoid} is not free of obstructions~\cite{boban2011impact}. GEMV2 therefore calculates whether 60\% of the radius of the first Fresnel ellipsoid is free of  obstructions to decide  whether the LOS path is~blocked.
%According to the performance evaluation in Sec.~\ref{sub:performance_results}, the LOS probability is a non-increasing function of the inter-vehicle distance  (the longer the link, the more likely to intersect  one or more blockages) and  depends on the considered propagation scenario, the type of blockage, and the density of the scenario.

 %such assumptions significantly undermine the accuracy of the analysis and might lead to wrong conclusions.

\vspace{-0.33cm}
\subsection{LOS Path Loss} % (fold)
\label{sub:los_path_loss}

As soon as the different communication states have been identified, the path loss follows a \emph{dual-slope piecewise-linear model}, which is deemed suitable to represent the real propagation in a vehicular environment. 
For \gls{los} transmissions, the  path loss  depends on the inter-vehicle distance $d$ and the considered scenario, and is computed as reported in Table~\ref{tab:PL}:
\medmuskip=3mu
%\thinmuskip=0mu
\thickmuskip=3mu
 \PLOS
\medmuskip=6mu
%\thinmuskip=0mu
\thickmuskip=6mu

In Eq.~\eqref{eq:PL_LOS}, $f_c$ is in GHz and $d$ is in meters. $\chi_a$ represents the shadowing, i.e., the effect of signal power fluctuations due to surrounding objects, and is  modeled according to a lognormal   random variable with standard deviation $\chi_{\sigma_a}=3$ dB~\cite{36885}.
For above 6 GHz propagation, oxygen absorption is modeled introducing additional loss which is derived based on~\cite{38901}, i.e., PL$_{\rm oxy} = d\cdot\Omega(f_c){}/1000$ dB. At 60~GHz, $\Omega(f_c)=15$ dB/km.\\ 

\vspace{-0.53cm}
\subsection{NLOSv/NLOS Path Loss} % (fold)
\label{sub:nlos_path_loss}

The 3GPP characterizes the NLOSv and NLOS path loss equations as illustrated in Table~\ref{tab:PL}, but does not distinguish between urban and highway~scenarios in case of NLOS.

        \emph{a) NLOSv:}
        The model  provides the additional attenuation factor $\mathcal{A}_{\rm NLOSv}$  to be summed to the LOS path loss. $\mathcal{A}_{\rm NLOSv}$ is modeled according to a lognormal random variable with mean $\mu_a$ and standard deviation $\sigma_a$. Different values of $\mu_a$ and $\sigma_a$ are obtained for different vehicle dropping strategies.

        \begin{itemize}
       	\item If the minimum between the antenna height of the TX and the RX is larger than the blocker height,  $\mu_a = \sigma_a=0$;
       	\item If the maximum between the antenna height  of the TX and the RX is smaller than the blocker~height,  $\mu_a = 9+\max(0,15\log_{10}(d)-41)$ dB and $\sigma_a=4.5$ dB;
       	\item For all the remaining configurations, $\mu_a = 5+\max(0,15\log_{10}(d)-41)$ dB and $\sigma_a=4$ dB.
       \end{itemize}

% \begin{table}[!t]
% \caption{Additional loss $\mathcal{A}^*_{\rm NLOSv}$, in case of NLOSv, as a function of  different nodes' dropping strategies.\vspace{0.33cm}}
% \label{tab:A_NLOSv_37885}
% \centering
% \footnotesize
% \begin{tabular}{c|c|c}
% \Xhline{2\arrayrulewidth}
%  \multicolumn{3}{c}{$\mathcal{A}_{\rm NLOSv}\sim \max\{0 \text{ dB, LogNormal}(\mu_a,\sigma_a)\}$}\rule[-0.2cm]{0pt}{0pt} \\\cline{1-3}
% Case A & Case B & Case C \\
% \Xhline{2\arrayrulewidth}
% \begin{tabular}[c]{@{}c@{}} $\mu_a = 0$\\ $\sigma_a=0$ \end{tabular} & \begin{tabular}[c]{@{}c@{}} $\mu_a = 12.5$ dB\\ $\sigma_a=4.5$ dB \end{tabular} & \begin{tabular}[c]{@{}c@{}} $\mu_a = 5$ dB\\ $\sigma_a=4$ dB \end{tabular} \\ 
% \Xhline{2\arrayrulewidth}
% \end{tabular}
% \end{table}

\begin{figure*}[b!]
     \centering
     \vspace{-0.33cm}
     \setlength{\belowcaptionskip}{-0.5cm}
\includegraphics[width=0.99\textwidth]{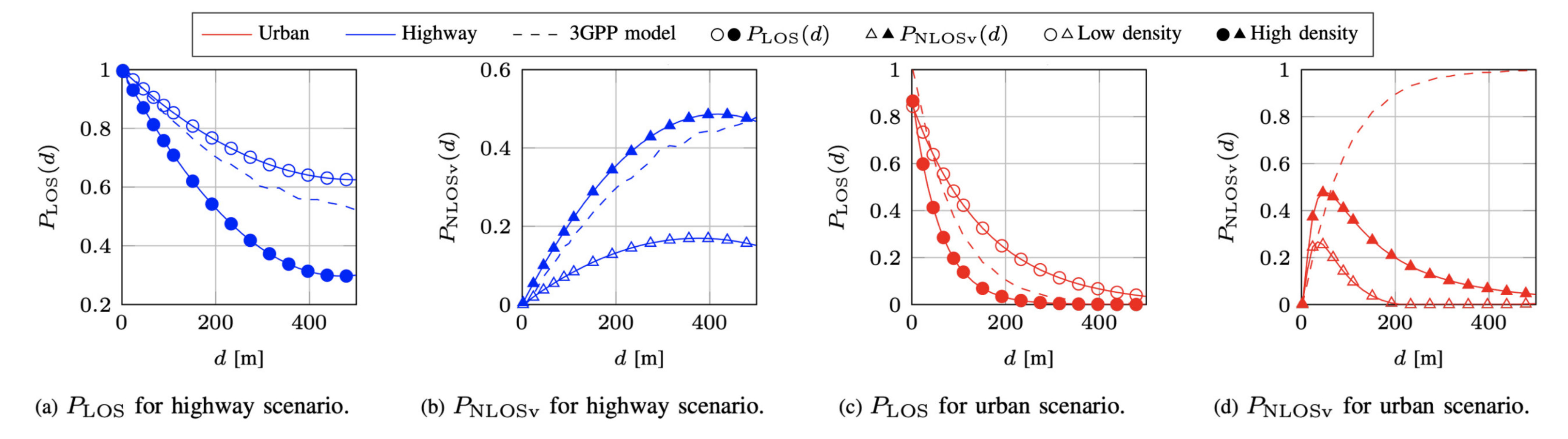}
\caption{$P_{\rm LOS}$ and $P_{\rm NLOSv}$ vs.  $d$ for urban and highway scenarios for different traffic densities. The 3GPP~\cite{37885} and the extended~\cite{boban2016modeling} models are~compared.}
\label{fig:model_comparison}
\end{figure*}

The NLOSv path loss  is finally computed as: 
\begin{equation}
\begin{cases}
	\text{PL}^{\rm h}_{\rm NLOSv}(d) = \text{PL}^{\rm h}_{\rm LOS}(d)+ \mathcal{A}_{\rm NLOSv }\\
	\text{PL}^{\rm u}_{\rm NLOSv}(d) = \text{PL}^{\rm u}_{\rm LOS}(d)+ \mathcal{A}_{\rm NLOSv}
	\label{eq:PL_NLOSv}
\end{cases}
\end{equation}

\smallskip

\emph{b) NLOS:}
        A frequency-dependent NLOS equation is defined (and valid for both urban and highway scenarios),~i.e.,
\medmuskip=0mu
%\thinmuskip=0mu
\thickmuskip=0mu
 \PNLOS
\medmuskip=6mu
%\thinmuskip=0mu
\thickmuskip=6mu
       where $f_c$ is in GHz, $d$ is in meters, and the shadowing component $\chi_a$ is modeled according to a lognormal   random variable with standard deviation $\chi_{\sigma_a}=4$ dB~\cite{36885}.
As mentioned, for above 6 GHz propagation, oxygen absorption is modeled based on~\cite{38901}, i.e., PL$_{\rm oxy} = d\cdot\Omega(f_c){}/1000$ dB.\\

\renewcommand{\arraystretch}{0.7}
\begin{table}[t!]
\footnotesize
\centering
\caption{Main system-level simulation parameters.}
\begin{tabularx}{0.95\columnwidth}{ @{\extracolsep{\fill}} lll}
\Xhline{2\arrayrulewidth}
\rule{0pt}{0.3cm}
Parameter & Value & Description \rule[-0.2cm]{0pt}{0pt} \\ \Xhline{2\arrayrulewidth}
$W_\ell$ & $\{3.5, 4\}$ m & \{Urban, highway\} lane width   \\
$N_\ell$ & $\{2, 3\}$ & \{Urban, highway\} number of lanes   \\ \midrule
$l_v$ & \{5, 13\} m & \{Type 2, Type 3\} vehicle length \\ 
$w_v$ & \{2, 2.6\} m & \{Type 2, Type 3\} vehicle  width \\ 
$h_v$ & \{1.6, 3\} m & \{Type 2, Type 3\} vehicle  height \\ 
$h_a$ & \{1.6, 3\} m & \{Type 2, Type 3\}   antenna height \\  \midrule
$P_{\rm TX}$ & $21$ dBm & Transmission power   \\
 $W_{\rm tot}$ & $1$ GHz & Total bandwidth \\
$f_{\rm c}$ & $63$ GHz &  Carrier frequency \\
NF & 13 dB & Noise figure\\
$N$ & $\{1,32\}$  & Antenna array size  \\
\Xhline{2\arrayrulewidth}
\end{tabularx}
\label{tab:params}
\vspace{-0.33cm}
\end{table}

\begin{figure*}[b!]
     \centering
        \setlength{\belowcaptionskip}{-0.5cm}
                     \begin{subfigure}[t!]{0.32\textwidth}
                     \centering
          \setlength{\belowcaptionskip}{0cm}
  \setlength{\belowcaptionskip}{0cm}
    \setlength\fwidth{0.99\columnwidth}
  \setlength\fheight{0.93\columnwidth}
\pgfplotsset{
tick label style={font=\scriptsize},
label style={font=\scriptsize},
legend  style={font=\scriptsize}
}
\begin{tikzpicture}
\begin{axis}[%
width=\fwidth,
height=\fheight,
at={(0\fwidth,0\fheight)},
xmin=0,
xmax=500,
ymin=70,
ymax=140,
axis background/.style={fill=white},
xmajorgrids,
ymajorgrids,
xlabel={$d$ [m]},
ylabel={LOS path loss [dB]},
legend columns={1},
label style={font=\scriptsize},
legend style={at={(0.65,0.1)},legend cell align=left, align=left, anchor = south, draw=white!15!black},
]

\addplot [color=blue, mark=*, mark size=2pt,  mark repeat=2,mark options={solid, fill=blue, blue}]
  table[row sep=crcr]{%
2	74.0136249209657\\
12.8888888888889	90.3606679367436\\
23.7777777777778	95.8431169525477\\
34.6666666666667	99.2812666992815\\
45.5555555555556	101.817185286637\\
56.4444444444444	103.842115731217\\
67.3333333333333	105.537627302224\\
78.2222222222222	107.002961335034\\
89.1111111111111	108.298328851237\\
100	109.463025007652\\
110	110.440878710833\\
130.473684210526	112.230588787128\\
150.947368421053	113.803746454863\\
171.421052631579	115.215623949401\\
191.894736842105	116.502707327618\\
212.368421052632	117.690350085576\\
232.842105263158	118.796886933501\\
253.315789473684	119.836007064285\\
273.789473684211	120.818202051095\\
294.263157894737	121.751690202184\\
314.736842105263	122.643029299954\\
335.210526315789	123.497535861079\\
355.684210526316	124.319579895279\\
376.157894736842	125.112797051674\\
396.631578947368	125.880244468549\\
417.105263157895	126.62451735379\\
437.578947368421	127.347837602559\\
458.052631578947	128.052122132759\\
478.526315789474	128.739036266483\\
499	129.410035920119\\
};
\addlegendentry{ LOS Highway}

\addplot [color=red, mark size=2pt, mark=o, mark repeat=2,mark options={solid, red}]
  table[row sep=crcr]{%
2	76.1895536845625\\
12.8888888888889	89.8662846027806\\
23.7777777777778	94.4710795309237\\
34.6666666666667	97.3688845694702\\
45.5555555555556	99.5133265899192\\
56.4444444444444	101.231093511131\\
67.3333333333333	102.673795672941\\
78.2222222222222	103.924299590377\\
89.1111111111111	105.032881466345\\
100	106.032352756965\\
110	106.873610599108\\
130.473684210526	108.418690881233\\
150.947368421053	109.782949902223\\
171.421052631579	111.012539978598\\
191.894736842105	112.137926967816\\
212.368421052632	113.180281039115\\
232.842105263158	114.154911675577\\
253.315789473684	115.073249353196\\
273.789473684211	115.94405453559\\
294.263157894737	116.774189510228\\
314.736842105263	117.569130025254\\
335.210526315789	118.333315372214\\
355.684210526316	119.070394509215\\
376.157894736842	119.78340320322\\
396.631578947368	120.474894164691\\
417.105263157895	121.147034392304\\
437.578947368421	121.801679168442\\
458.052631578947	122.440429119602\\
478.526315789474	123.064674789691\\
499	123.675631868886\\
};
\addlegendentry{LOS Urban}

\end{axis}
% \spy [blue, size=3cm] on (spypoint)
%    in node[fill=white] at (magnifyglass);
\end{tikzpicture}%
    \caption{\footnotesize  LOS path loss. Type 2 vehicles are deployed.}
    \label{fig:PL_LOS}
      \end{subfigure} \,
                   \begin{subfigure}[t!]{0.32\textwidth}
                   \centering
                \setlength{\belowcaptionskip}{0cm}
  \setlength{\belowcaptionskip}{0cm}
    \setlength\fwidth{0.99\columnwidth}
  \setlength\fheight{0.93\columnwidth}
  \usetikzlibrary{spy}
\definecolor{mycolor1}{rgb}{1.00000,0.00000,1.00000}%
\pgfplotsset{
tick label style={font=\scriptsize},
label style={font=\scriptsize},
legend  style={font=\scriptsize}
}
\begin{tikzpicture}[spy using outlines=
	{circle, magnification=2, connect spies}]

\begin{axis}[%
width=\fwidth,
height=\fheight,
at={(0\fwidth,0\fheight)},
xmin=0,
xmax=500,
ymin=40,
ymax=170,
axis background/.style={fill=white},
xmajorgrids,
ymajorgrids,
xlabel style={font=\scriptsize\color{white!15!black}},
ylabel style={font=\scriptsize\color{white!15!black}},
xlabel={$d$ [m]},
ylabel={NLOSv/NLOS path loss [dB]},
legend columns={1},
label style={font=\scriptsize},
legend style={at={(0.51,0.52)},legend cell align=left, align=left, anchor = north, draw=white!15!black},
]
\addplot [color=blue, mark=triangle*, mark size = 2.5pt,mark repeat=2, mark options={solid, fill=blue, blue}]
  table[row sep=crcr]{%
2	79.099583006015\\
12.8888888888889	95.296533079816\\
23.7777777777778	100.933776931533\\
34.6666666666667	104.224206449365\\
45.5555555555556	106.961431525649\\
56.4444444444444	109.025839600665\\
67.3333333333333	110.366243615142\\
78.2222222222222	112.091939544302\\
89.1111111111111	113.316628587258\\
100	114.369245798166\\
110	115.153793299078\\
130.473684210526	117.109342430148\\
150.947368421053	118.6879687267\\
171.421052631579	120.563656266975\\
191.894736842105	121.434610246348\\
212.368421052632	122.959522135346\\
232.842105263158	123.654464737945\\
253.315789473684	124.910825794862\\
273.789473684211	125.965179593314\\
294.263157894737	126.82735532928\\
314.736842105263	127.637450638681\\
335.210526315789	128.710881196174\\
355.684210526316	129.295465592255\\
376.157894736842	130.237071031868\\
396.631578947368	130.84261454157\\
417.105263157895	131.608757388091\\
437.578947368421	132.48612764496\\
458.052631578947	132.908407174088\\
478.526315789474	133.674405352762\\
499	134.528381352341\\
};
\addlegendentry{NLOSv Highway (Type 2)}

\addplot [color=blue, dashed]
  table[row sep=crcr]{%
2	78.2728452848261\\
12.8888888888889	94.8276914660609\\
23.7777777777778	100.264435810494\\
34.6666666666667	103.346893141359\\
45.5555555555556	106.073255991049\\
56.4444444444444	108.2928393362\\
67.3333333333333	110.006545893105\\
78.2222222222222	111.285749529458\\
89.1111111111111	112.642976280271\\
100	113.631752716238\\
110	114.720241867558\\
130.473684210526	116.672261705285\\
150.947368421053	117.966742586677\\
171.421052631579	119.479492618842\\
191.894736842105	120.937620393649\\
212.368421052632	121.937386237694\\
232.842105263158	122.901407247422\\
253.315789473684	124.155931464424\\
273.789473684211	125.347120862383\\
294.263157894737	125.858208551494\\
314.736842105263	126.883663624519\\
335.210526315789	127.697827194535\\
355.684210526316	128.51992932723\\
376.157894736842	129.319043306484\\
396.631578947368	130.279933512324\\
417.105263157895	130.736165771998\\
437.578947368421	131.680436694729\\
458.052631578947	132.465328732728\\
478.526315789474	133.045229721632\\
499	133.787127181547\\
};
\addlegendentry{NLOSv Highway (Type 2+3)}

\addplot [color=red, ,  mark size = 2.5pt,  mark=triangle, mark repeat=2,mark options={solid, red, fill=red}]
  table[row sep=crcr]{%
2	81.3502338111708\\
12.8888888888889	95.0202107109498\\
23.7777777777778	99.4871811494166\\
34.6666666666667	102.487215579099\\
45.5555555555556	104.623172084983\\
56.4444444444444	106.151952877579\\
67.3333333333333	107.801986844424\\
78.2222222222222	109.110065326868\\
89.1111111111111	110.077466117998\\
100	111.116676479365\\
110	111.835748232257\\
130.473684210526	113.330743587974\\
150.947368421053	115.061701890651\\
171.421052631579	116.044975946891\\
191.894736842105	117.238862359687\\
212.368421052632	118.14426550947\\
232.842105263158	119.123687856738\\
253.315789473684	119.856660384599\\
273.789473684211	120.935437015448\\
294.263157894737	121.611543535759\\
314.736842105263	122.555221120593\\
335.210526315789	123.425624462797\\
355.684210526316	124.095282263239\\
376.157894736842	124.963561759719\\
396.631578947368	125.536188767048\\
417.105263157895	126.156128518984\\
437.578947368421	126.967116888274\\
458.052631578947	127.487761729497\\
478.526315789474	128.085145283248\\
499	128.839911781658\\
};
\addlegendentry{NLOSv Urban (Type 2)}

\addplot [color=red, dashed]
  table[row sep=crcr]{%
2	80.3999316860438\\
12.8888888888889	94.1221191103703\\
23.7777777777778	98.8849509213903\\
34.6666666666667	101.469919297937\\
45.5555555555556	103.785430022519\\
56.4444444444444	105.471282826896\\
67.3333333333333	106.793129632558\\
78.2222222222222	108.194229451937\\
89.1111111111111	109.165494170556\\
100	110.327216157165\\
110	111.355772076544\\
130.473684210526	112.875967563979\\
150.947368421053	114.29831592415\\
171.421052631579	115.167974619958\\
191.894736842105	116.593481899094\\
212.368421052632	117.535471064272\\
232.842105263158	118.700907191131\\
253.315789473684	119.390673522762\\
273.789473684211	120.113409224283\\
294.263157894737	121.041066855895\\
314.736842105263	121.949429327856\\
335.210526315789	122.869393220661\\
355.684210526316	123.424771100172\\
376.157894736842	124.498440160267\\
396.631578947368	124.837873573058\\
417.105263157895	125.303808632112\\
437.578947368421	125.949564987948\\
458.052631578947	126.686658920365\\
478.526315789474	127.465253110538\\
499	127.900229188475\\
};
\addlegendentry{NLOSv Urban (Type 2+3)}

\addplot [color=black,  mark size = 1.5pt, mark=square, mark repeat=2,mark options={solid, black}]
  table[row sep=crcr]{%
2	79.5179585021688\\
12.8888888888889	103.956856359234\\
23.7777777777778	112.09886321622\\
34.6666666666667	117.174421169641\\
45.5555555555556	120.896632384019\\
56.4444444444444	123.852361384289\\
67.3333333333333	126.313962074069\\
78.2222222222222	128.430296456674\\
89.1111111111111	130.291681064275\\
100	131.957058632263\\
110	133.348839186994\\
130.473684210526	135.879851669845\\
150.947368421053	138.086035539861\\
171.421052631579	140.0502991501\\
191.894736842105	141.827371585877\\
212.368421052632	143.455283091176\\
232.842105263158	144.961535731543\\
253.315789473684	146.366663296076\\
273.789473684211	147.686403144698\\
294.263157894737	148.933082739827\\
314.736842105263	150.116538754897\\
335.210526315789	151.244745964959\\
355.684210526316	152.324259384752\\
376.157894736842	153.360532487744\\
396.631578947368	154.358150981386\\
417.105263157895	155.321007677746\\
437.578947368421	156.252435419262\\
458.052631578947	157.155309583045\\
478.526315789474	158.032128152128\\
499	158.885075000974\\
};
\addlegendentry{NLOS Urban/Highway}

\coordinate (spypoint) at (axis cs:200,120);
  \coordinate (magnifyglass) at (axis cs:85,160);

\end{axis}

\spy [blue, size=1.3cm] on (spypoint)
  in node[fill=white] at (magnifyglass);
\end{tikzpicture}%
    \caption{\footnotesize NLOSv and NLOS path loss.}
    \label{fig:PL_NLOS}
      \end{subfigure}\,
      \begin{subfigure}[t!]{0.32\textwidth}
     \centering
        \setlength{\belowcaptionskip}{-0.5cm}
          \setlength{\belowcaptionskip}{0cm}
  \setlength{\belowcaptionskip}{0cm}
    \setlength\fwidth{0.99\columnwidth}
  \setlength\fheight{0.93\columnwidth}
\usetikzlibrary{spy}
\definecolor{mycolor1}{rgb}{1.00000,0.00000,1.00000}%
\pgfplotsset{
tick label style={font=\scriptsize},
label style={font=\scriptsize},
legend  style={font=\scriptsize}
}
\begin{tikzpicture}[spy using outlines=
	{circle, magnification=2,connect spies}]

\begin{axis}[%
width=\fwidth,
height=\fheight,
at={(0\fwidth,0\fheight)},
xmin=0,
xmax=500,
ymin=70,
ymax=160,
axis background/.style={fill=white},
xlabel={$d$ [m]},
ylabel={Overall path loss [dB]},
xmajorgrids,
ymajorgrids,
legend columns={1},
label style={font=\scriptsize},
legend style={at={(0.58,0.42)},legend cell align=left, align=left, anchor = north, draw=white!15!black},
]
\addplot [color=blue, mark=diamond*, mark size = 2pt, mark repeat=2,mark options={solid, blue}]
  table[row sep=crcr]{%
2	74.0341621692749\\
12.8888888888889	90.5607896316313\\
23.7777777777778	96.177560054022\\
34.6666666666667	99.8319138508173\\
45.5555555555556	102.598875119729\\
56.4444444444444	104.774501027527\\
67.3333333333333	106.678099463714\\
78.2222222222222	108.3737809956\\
89.1111111111111	109.793772533949\\
100	111.133910639028\\
110	112.407883096166\\
130.473684210526	114.483203142181\\
150.947368421053	116.371455909583\\
171.421052631579	118.080182917527\\
191.894736842105	119.834759940878\\
212.368421052632	121.425934444153\\
232.842105263158	122.797252289064\\
253.315789473684	124.090181426547\\
273.789473684211	125.469122876579\\
294.263157894737	126.633131883633\\
314.736842105263	127.810394607333\\
335.210526315789	128.928678462112\\
355.684210526316	129.993142066814\\
376.157894736842	131.111819560849\\
396.631578947368	132.294916897312\\
417.105263157895	133.108336750534\\
437.578947368421	134.201561155548\\
458.052631578947	134.844866925585\\
478.526315789474	135.997018484396\\
499	136.870155201947\\
};
\addlegendentry{Highway  (Low density)}

\addplot [color=blue, dashed]
  table[row sep=crcr]{%
2	74.0398381574834\\
12.8888888888889	90.6292017909249\\
23.7777777777778	96.3621715538078\\
34.6666666666667	100.133607598056\\
45.5555555555556	102.898471823317\\
56.4444444444444	105.256497800653\\
67.3333333333333	107.124409624569\\
78.2222222222222	108.854215838752\\
89.1111111111111	110.514855325997\\
100	111.892300618801\\
110	113.129801048694\\
130.473684210526	115.495184146308\\
150.947368421053	117.424887057503\\
171.421052631579	119.216459312812\\
191.894736842105	120.984437255301\\
212.368421052632	122.558342260015\\
232.842105263158	124.366892090188\\
253.315789473684	125.689944610389\\
273.789473684211	127.21572622795\\
294.263157894737	128.481663247355\\
314.736842105263	129.598505900835\\
335.210526315789	130.846908807442\\
355.684210526316	132.064789013129\\
376.157894736842	133.017362869665\\
396.631578947368	133.865065252947\\
417.105263157895	135.084640804373\\
437.578947368421	135.724486512039\\
458.052631578947	137.085391454541\\
478.526315789474	137.677455953803\\
499	138.394148771079\\
};
\addlegendentry{Highway (High density)}

\addplot [color=red, mark=diamond, mark size = 2pt,mark repeat=2, mark options={solid, red, fill=red}]
  table[row sep=crcr]{%
2	76.7550496630737\\
12.8888888888889	91.4738201996376\\
23.7777777777778	96.0950835124828\\
34.6666666666667	99.7016646235126\\
45.5555555555556	103.129054989975\\
56.4444444444444	106.238378464351\\
67.3333333333333	109.483467812643\\
78.2222222222222	112.351545977723\\
89.1111111111111	115.124734653586\\
100	117.838366792951\\
110	119.924829337527\\
130.473684210526	124.364401392418\\
150.947368421053	128.050596032651\\
171.421052631579	131.414907650254\\
191.894736842105	134.389397437753\\
212.368421052632	136.825057641774\\
232.842105263158	138.957324703032\\
253.315789473684	141.22515538526\\
273.789473684211	142.896482739586\\
294.263157894737	144.652734050966\\
314.736842105263	146.474483718052\\
335.210526315789	147.950311762627\\
355.684210526316	149.587466305495\\
376.157894736842	151.016848863683\\
396.631578947368	152.040536215123\\
417.105263157895	153.318412843217\\
437.578947368421	154.371424127967\\
458.052631578947	155.501757970316\\
478.526315789474	156.649517849902\\
499	157.627950138242\\
};
\addlegendentry{Urban (Low density)}

\addplot [color=red, dashed]
  table[row sep=crcr]{%
2	76.6594520414298\\
12.8888888888889	91.9259023950711\\
23.7777777777778	96.8501760445007\\
34.6666666666667	100.667410166227\\
45.5555555555556	104.216753462567\\
56.4444444444444	107.754177777708\\
67.3333333333333	111.185876017708\\
78.2222222222222	114.146581889111\\
89.1111111111111	117.069160469705\\
100	119.719326953908\\
110	121.955913166336\\
130.473684210526	126.30980186934\\
150.947368421053	129.723664794561\\
171.421052631579	132.836315852333\\
191.894736842105	135.693386224093\\
212.368421052632	138.133721887572\\
232.842105263158	140.246762370256\\
253.315789473684	142.015423165045\\
273.789473684211	144.054286601043\\
294.263157894737	145.541687093722\\
314.736842105263	147.066076554443\\
335.210526315789	148.545105602036\\
355.684210526316	150.062363614555\\
376.157894736842	151.11719201938\\
396.631578947368	152.254083307862\\
417.105263157895	153.448925954607\\
437.578947368421	154.659668817938\\
458.052631578947	155.619081025391\\
478.526315789474	156.536540562146\\
499	157.599615084888\\
};
\addlegendentry{Urban (High density)}
\coordinate (spypoint) at (axis cs:200,135);
  \coordinate (magnifyglass) at (axis cs:100,150);

\end{axis}

\spy [blue,size=1.3cm] on (spypoint)
   in node[fill=white] at (magnifyglass);
\end{tikzpicture}%
\caption{Overall path loss. Type 2 vehicles are deployed.}
\label{fig:PL_all}
\end{subfigure}
\caption{Path loss vs.  $d$ for urban and highway scenarios for different traffic densities and deployment options. The extended model in \cite{boban2016modeling} is~considered.}
\label{fig:Pathloss}
\end{figure*}
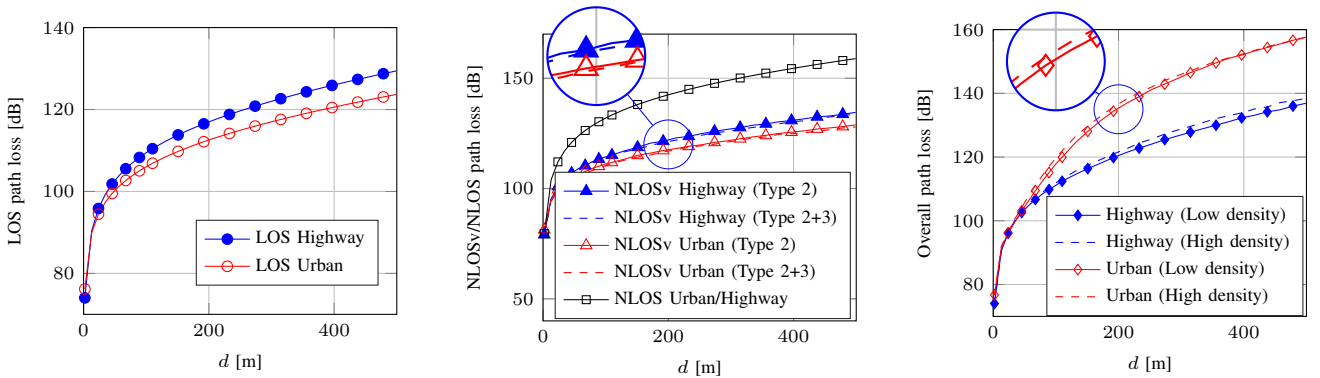

\vspace{-0.5cm}
\section{Performance Evaluation} % (fold)
\label{sec:performance_evaluation}
In this section, we  review the  methodology defined in \cite{37885} for assessing the performance of  vehicular~networks,  and 
 validate the 3GPP V2V path loss model proposed for nodes operating at mmWaves through system-level simulations.
\vspace{-0.33cm}
\subsection{Evaluation Methodology} % (fold)
\label{sub:evaluation_methodology}

\emph{a) Evaluation Scenarios.} 
The channel characteristics are heavily influenced by the proprieties of the environment in which the vehicles are deployed, i.e., urban or highway.
The parameters regarding the road configuration for both scenarios are taken from~\cite[Table A-1]{37885}:
for the urban (highway) case, $N_\ell=2$ ($N_\ell=3$) lanes per directions are assumed, where each lane has width $W_\ell=3.5$ m ($W_\ell=4$ m).
\medskip

% \begin{figure*}[b!]
%      \centering
%         \setlength{\belowcaptionskip}{-0.5cm}
%                      \begin{subfigure}[t!]{0.32\textwidth}
%                      \centering
%           \setlength{\belowcaptionskip}{0cm}
%   \setlength{\belowcaptionskip}{0cm}
%   \setlength\fwidth{0.99\columnwidth}
%   \setlength\fheight{0.99\columnwidth}
%   \input{Figures/PL_LOS.tex}
%     \caption{\footnotesize  LOS path loss. Type 2 vehicles are deployed.}
%     \label{fig:PL_LOS}
%       \end{subfigure} \,
%                    \begin{subfigure}[t!]{0.32\textwidth}
%                    \centering
%                 \setlength{\belowcaptionskip}{0cm}
%   \setlength{\belowcaptionskip}{0cm}
%   \setlength\fwidth{0.99\columnwidth}
%   \setlength\fheight{0.99\columnwidth}
%   \input{Figures/PL_NLOSv.tex}
%     \caption{\footnotesize NLOSv and NLOS path loss.}
%     \label{fig:PL_NLOS}
%       \end{subfigure}\,
%       \begin{subfigure}[t!]{0.32\textwidth}
%      \centering
%         \setlength{\belowcaptionskip}{-0.5cm}
%           \setlength{\belowcaptionskip}{0cm}
%   \setlength{\belowcaptionskip}{0cm}
%   \setlength\fwidth{0.99\columnwidth}
%   \setlength\fheight{0.99\columnwidth}
%   \input{Figures/PL_37885_density.tex}
% \caption{Overall path loss. Type 2 vehicles are deployed.}
% \label{fig:PL_all}
% \end{subfigure}
% \caption{Path loss vs. inter-vehicle distance $d$ for urban and highway scenarios. Different densities of vehicular traffic and deployment options are considered. }
% \label{fig:Pathloss}
% \end{figure*}

\emph{b) Vehicle Characteristics.}
Three types of vehicles are defined according to the 3GPP~\cite[Sec. 6.1.2]{37885}: Type 1 and Type 2 vehicles are passenger cars with   bumper and rooftop antenna position, respectively, while Type 3 vehicles are large trucks or buses. %The model specifies the vehicle length, width and height,  and the antenna height. %\footnote{Although not specified in \cite{37885}, the dimension of the  vehicle widths and heights can also be statistically modeled as normal random variables. Detailed parameters (i.e., mean and std. deviation of vehicle widths and heights) for passenger cars and trucks can be found in~\cite{boban2011impact}.}
The vehicle type may potentially  affect the path loss equation (since the antenna height is determined according to the type of vehicle), the loss caused by the blocking vehicle, and  the radiation pattern. 
In our analysis we   consider both Type 2 and Type 3 deployments. 
\medskip

\emph{c) Vehicle Dropping Models.}
Vehicles are dropped according to a random process, so that the distance  between the rear bumper of a vehicle and the front bumper of the following vehicle in the same lane is equal to  $\max\{2 \text{ m}, \text{Exp}(\lambda)\}$, where $\lambda =\bar{v}\cdot 2 \text{ s.}$ and $\bar{v}$ is the average speed~\cite[Sec. 6.1.2]{37885}.
\medskip

\emph{d) System-Level Simulation Parameters.}
The simulation parameters  are based on the system-design considerations specified in~\cite[Sec. 6.1.1]{37885} and are summarized in Table~\ref{tab:params}.
For \gls{mmwave} links, the central frequency is set to  63~GHz while the total bandwidth is set to  1~GHz. %  for sidelink transmissions.
The vehicles' noise figure is set to 13 dB and the  transmit power is set to 21 dBm. % (the overall Equivalent Isotropic Radiated Power (EIRP) should not exceed 43 dBm~though).
In order to establish directional transmissions, vehicles are equipped with  Uniform Planar Arrays (UPAs) of $N$ elements. %\footnote{We assume that four equal and independent antenna arrays are placed on either side of each vehicle, so that each array is responsible for covering a $\Delta_\theta = 90^\circ$  azimuth space on the same elevation plane~\cite{giordani2018feasibility}.} 
For above 6 GHz propagation, the maximum value of $N$ is currently set to 32~\cite[Tab. 6.1.4-12]{37885}.
For completeness, in our study we also consider omnidirectional \gls{mmwave} transmissions, i.e., $N=1$.
Our results are finally derived through a Monte Carlo approach as a function of the inter-vehicle distance $d$, with $d$ varying from 2 m to 500 m. 
\vspace{-0.33cm}
\subsection{Performance Results} % (fold)
\label{sub:performance_results}

\textbf{Path Loss Probabilities.} 
In Fig.~\ref{fig:model_comparison} we plot the LOS and NLOSv probabilities vs. $d$ considering both the 3GPP model~\cite{37885} (which does not distinguish between low- and high-traffic densities) and the extended model~\cite{boban2016modeling}.
As foreseen, the LOS probability is significantly higher in case of highway deployments than in urban scenarios since the signal usually propagates in free space.
Moreover, it is clear that the impact of different density regimes is not negligible (the gap is particularly evident for large distances).
While, for the NLOSv case, the 3GPP model behaves as in a high-density scenario, for the LOS case it operates as in a low-density scenario, thereby setting a lower bound to the path loss. 
%In fact the additional NLOS state introduced by the model in~\cite{boban2016modeling} makes the NLOSv state less likely than it is following the 3GPP characterization.
For the urban case, $P_{\rm NLOSv}$ peaks at around $d=50$ m and then starts decreasing for larger values of $d$ when the model in~\cite{boban2016modeling} is considered. In fact, although the probability of both dynamic and static blockages potentially obstructing the propagation path between the endpoints increases with $d$,~\cite{boban2016modeling} assumes that the channel condition is categorized as NLOS when the line of sight is blocked by both vehicles and buildings.
The 3GPP model~\cite{37885}, instead, does not make this distinction and shows a monotonically increasing trend.

\textbf{Path Loss.}
The following results are derived considering the extended model in~\cite{boban2016modeling}.
The 3GPP model, in fact, does not define a closed-form expression for the NLOS probability and prevents a complete stochastic analysis for the path loss, which has to be based on geometric simulations instead.
In Fig. \ref{fig:Pathloss} we plot the path loss as a function of  $d$. Different densities of vehicular traffic and vehicle deployment options are considered.
We see that, for the LOS case (Fig. \ref{fig:Pathloss}a),  the 3GPP model registers better propagation in urban rather than highway scenario (i.e., around 5 dB at 200 m).
In fact, while in the highway environment  the
propagating signals attenuate over distance following Friis' law, 
in the urban environment the observed path loss is significantly lower,  indicating a waveguide effect resulting from the more likely reflections from walls of static blockages in street canyons. 

From Fig.~\ref{fig:Pathloss}b, we observe that, for the NLOSv case, the path loss slightly decreases when deploying both Type 2 and Type 3  vehicles, i.e., when tall vehicles (e.g., trucks) are deployed. 
In fact, although Type 3 blockage implies higher attenuation, larger vehicle heights may guarantee higher LOS probability when the obstacle is small. 

Moreover, we see that the NLOS path loss is generally more than 20 dB higher than its NLOSv counterpart, demonstrating the much stronger impact of static/environmental blockages like buildings or vegetation, compared to dynamic obstructions like pedestrians and cars, on the received signal strength.
We recall that  the 3GPP model does not distinguish between urban and highway propagation for the NLOS case.

Finally, Fig.~\ref{fig:Pathloss}c measures the overall path loss as a function of  $d$.
We observe that the urban path loss is significantly higher than its highway counterpart (although the waveguide effect caused by the more likely signal reflections and scattering   in street canyons generally results in reduced attenuation) due to the much higher probability of blockage intersection in contrast to free-space  propagation.
Furthermore, the higher the vehicle density, the more probable the NLOSv state and, therefore, the larger the overall sidelink path loss.

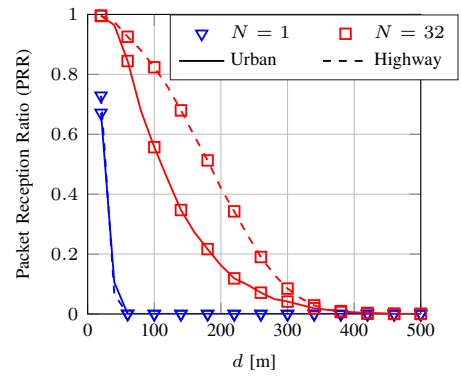
\begin{figure}[t!]
     \centering
        \setlength{\belowcaptionskip}{-0.5cm}
  \setlength\fwidth{0.5\columnwidth}
  \setlength\fheight{0.45\columnwidth}
\pgfplotsset{
tick label style={font=\scriptsize},
label style={font=\scriptsize},
legend  style={font=\scriptsize}
}
\begin{tikzpicture}

\begin{axis}[%
every axis plot/.append style={ line width=0.7pt},
width=\fwidth,
height=\fheight,
at={(0\fwidth,0\fheight)},
scale only axis,
xmin=0,
xmax=25,
xtick={0,5,10,15,20,25},
xticklabels={0,100,200,300,400,500},
ymin=0,
ymax=1,
xlabel={$d$ [m]},
ylabel={Packet Reception Ratio (PRR)},
yminorticks=true,
axis background/.style={fill=white},
xmajorgrids,
ymajorgrids,
yminorgrids,
legend columns = {2},
legend style={at={(0.68,0.78)}, anchor=south, legend cell align=left, align=left, draw=white!15!black,/tikz/every even column/.append style={column sep=0.35cm}}
]
\addplot [color=blue, dashed,  mark=triangle,mark size=2.5pt, mark repeat=2, mark options={solid, rotate=180, blue},forget plot]
  table[row sep=crcr]{%
1	0.728548208951478\\
2	0.0650601181984918\\
3	0\\
4	0\\
5	0\\
6	0\\
7	0\\
8	0\\
9	0\\
10	0\\
11	0\\
12	0\\
13	0\\
14	0\\
15	0\\
16	0\\
17	0\\
18	0\\
19	0\\
20	0\\
21	0\\
22	0\\
23	0\\
24	0\\
25	0\\
26	0\\
};
%\addlegendentry{data1}

% \addplot [color=black, dashed, mark=o,mark size=2pt, mark repeat=2, mark options={solid, black},forget plot]
%   table[row sep=crcr]{%
% 1	0.997464753141284\\
% 2	0.971010677129156\\
% 3	0.919357765646498\\
% 4	0.866012192617228\\
% 5	0.799281376341855\\
% 6	0.713923472404414\\
% 7	0.614775153393734\\
% 8	0.510865865959354\\
% 9	0.403536105451955\\
% 10	0.284856251876893\\
% 11	0.193848973087153\\
% 12	0.119822422706035\\
% 13	0.0624137333836698\\
% 14	0.0267004144023427\\
% 15	0.00954885064951968\\
% 16	0.00240206753601134\\
% 17	0.000384993946362156\\
% 18	2.60260781302865e-05\\
% 19	1.30548302872063e-05\\
% 20	0\\
% 21	0\\
% 22	0\\
% 23	0\\
% 24	0\\
% 25	0\\
% 26	0\\
% };
%\addlegendentry{data2}

\addplot [color=red, dashed, mark=square,mark size=2pt, mark repeat=2, mark options={solid, red},forget plot]
  table[row sep=crcr]{%
1	0.996205748335086\\
2	0.975327009702011\\
3	0.925331818238737\\
4	0.880127800665608\\
5	0.823621937380241\\
6	0.75806993147167\\
7	0.679707827935083\\
8	0.594821864415221\\
9	0.512659203374706\\
10	0.419842713592715\\
11	0.342527722561853\\
12	0.26304227265363\\
13	0.190208988164983\\
14	0.131147149494161\\
15	0.0848075378667613\\
16	0.0523769063180827\\
17	0.0287436465528477\\
18	0.0147908387813528\\
19	0.00629624793244537\\
20	0.00261304994314703\\
21	0.000855645091413222\\
22	0.000229568411386593\\
23	0\\
24	0\\
25	0\\
26	0\\
};
%\addlegendentry{data4}

\addplot [color=blue, mark=triangle,mark size=2.5pt, mark repeat=2, mark options={solid, rotate=180, blue},forget plot]
  table[row sep=crcr]{%
1	0.671308048661067\\
2	0.105588083079245\\
3	0.000844572476103182\\
4	0\\
5	0\\
6	0\\
7	0\\
8	0\\
9	0\\
10	0\\
11	0\\
12	0\\
13	0\\
14	0\\
15	0\\
16	0\\
17	0\\
18	0\\
19	0\\
20	0\\
21	0\\
22	0\\
23	0\\
24	0\\
25	0\\
26	0\\
};
%\addlegendentry{data5}

% \addplot [color=black, mark=o,mark size=2pt, mark repeat=2, mark options={solid, black},forget plot,forget plot]
%   table[row sep=crcr]{%
% 1	0.995110622173328\\
% 2	0.958221760239884\\
% 3	0.828329100280477\\
% 4	0.679219233026447\\
% 5	0.53981018437313\\
% 6	0.423572752946272\\
% 7	0.332029023900943\\
% 8	0.250907990314769\\
% 9	0.189990242728435\\
% 10	0.140697060004648\\
% 11	0.101514899195914\\
% 12	0.0730588958088704\\
% 13	0.0496307577314767\\
% 14	0.0331014046263903\\
% 15	0.0206089268443871\\
% 16	0.0124441891433196\\
% 17	0.00783615842407094\\
% 18	0.00372556684910086\\
% 19	0.00224894187454919\\
% 20	0.00115669387933021\\
% 21	0.00028341418461913\\
% 22	0.000170227549070697\\
% 23	3.48647682800338e-05\\
% 24	0\\
% 25	0\\
% 26	0\\
% };
%\addlegendentry{data6}

\addplot [color=red, mark=square,mark size=2pt, mark repeat=2, mark options={solid, red},forget plot]
  table[row sep=crcr]{%
1	0.994791802071147\\
2	0.964311122679161\\
3	0.84442560034171\\
4	0.677702660289004\\
5	0.556885134758623\\
6	0.449810303637963\\
7	0.347134669591686\\
8	0.273386839985764\\
9	0.21661436185029\\
10	0.162358987760791\\
11	0.118871584496585\\
12	0.0973706031247274\\
13	0.071124970478925\\
14	0.0499320962504783\\
15	0.0414776409591704\\
16	0.030296330991413\\
17	0.01908014571949\\
18	0.0138098948188547\\
19	0.00975305455171227\\
20	0.00571023965141612\\
21	0.00355935922756199\\
22	0.00237254901960784\\
23	0.000981033355134074\\
24	0.000185476128131273\\
25	0.000471903270702853\\
26	0.000216590859865714\\
};
%\addlegendentry{data8}

\addplot [color=blue, only marks, mark=triangle,mark size=2.5pt, mark repeat=2, mark repeat=2, mark options={solid, rotate=180, blue}]
  table[row sep=crcr]{%
-1	0.00001\\
-2 0.00001 \\
};
\addlegendentry{$N=1$}

% \addplot [color=red, only marks, mark=o,mark size=2pt, mark repeat=2,mark size=2pt, mark repeat=2, mark options={solid, rotate=180, red}]
%   table[row sep=crcr]{%
% -1	0.00001\\
% -2 0.00001 \\
% };
% \addlegendentry{$N=16$}

\addplot [color=red, only marks, mark=square,mark size=2pt, mark repeat=2, mark repeat=2, mark options={solid, rotate=180, red}]
  table[row sep=crcr]{%
-1	0.00001\\
-2 0.00001 \\
};
\addlegendentry{$N=32$}

\addplot [color=black]
  table[row sep=crcr]{%
-1	0.00001\\
-2 0.00001 \\
};
\addlegendentry{Urban}

\addplot [color=black,dashed]
  table[row sep=crcr]{%
-1	0.00001\\
-2 0.00001 \\
};
\addlegendentry{Highway}

\end{axis}
\end{tikzpicture}%
    \caption{\footnotesize  PRR vs. inter-vehicle distance $d$ for urban and highway scenarios in a medium traffic density environment and for different array configurations. }
    \label{fig:PRR}
\end{figure}

\textbf{Packet Reception Ratio.}
The performance of the vehicular network can be assessed in terms of average \emph{\gls{prr}},~\cite[Sec. 6.1.6]{37885}, which measures the percentage of vehicles experiencing successful packet reception.
In this paper, successful reception is achieved if the \gls{snr} experienced between the transmitting and receiving vehicles   is above a predefined threshold, taken to be $0$ dB in our simulations.
%High values of \gls{prr} usually ensure more reliable V2V communications, a critical prerequisite for safety services requiring ubiquitous and continuous connectivity.
Fig.~\ref{fig:PRR}  plots the \gls{prr} vs. $d$, for different antenna array configurations. 
We observe that higher \gls{prr} 
is guaranteed when considering short-range communications, since the endpoints are progressively
closer, thus ensuring 
stronger received power. 
Moreover, we see that  the \gls{prr} is unacceptably low (PRR $< 0.8$) in case of omnidirectional transmissions (i.e., $N = 1$) for both urban and highway scenarios, thereby exemplifying how single-antenna communications should be avoided in a mmWave  context.
% (i.e., up to a few tens of meters) which therefore represents a suitable range for omnidirectional mmWave links in vehicular scenarios).
On the other hand, the \gls{prr} increases in case of directional transmissions, thanks to the high gains  produced by beamforming.
The benefits are particularly evident considering highway propagation, thanks to the much lower path loss that is generally experienced. 
%According to the 3GPP specifications~\cite{3GPP_22186}, high values of \gls{prr} (e.g., $> 99\%$ for advanced driving applications) are recommended to ensure reliable V2V communications, a critical concern for safety services requiring   continuous and ubiquitous connectivity.
%If no retransmissions are considered, such requirements can be achieved by employing very large antenna arrays and through  dedicated close-range transmissions, i.e., up to a few tens of meters, which therefore represents a suitable range for mmWave links in vehicular scenarios.

\vspace{-0.33cm}
\section{Open Challenges} % (fold)
\label{sec:open_challenges}
The \gls{mmwave} bands hold great promise to support cooperative driving applications because of the large available bandwidth which may provide the required link capacity.
Before unleashing the potential of this technology, it is  important to  demonstrate the feasibility of designing \gls{mmwave}-aware strategies in a vehicular context.
%While traditional ray-tracing measurement campaigns have shown strong dependency on the accuracy of the data used to describe the environment,  probabilistic models may be preferred to ensure statistically reliable performance~analyses.
In this regard, the 3GPP has recently proposed a model for the V2V channel at mmWaves whose accuracy, however, has not yet been fully examined. 
Motivated by these considerations, in this paper we  provided the first numerical validation of the 3GPP V2V path loss model for nodes operating at mmWaves considering realistic system-level simulation parameters.

Although the results presented in the above sections provide some valuable insights into the propagation characteristics of the mmWave signals in a vehicular scenario, there remain many open problems which call for innovative modeling and design solutions.
A list of relevant challenges includes the following aspects. 
\begin{itemize}

  \item \emph{Propagation Scenario. }
  It has been observed that the distance dependency of the  path loss  has remarkable sensitivity to the surrounding environment.
  In this regard, the 3GPP distinguishes between urban and highway propagation only in case of LOS transmissions, and does not provide a closed-form expression for the NLOS probability. 
  A broader classification of the different propagation scenarios might be necessary.
  %In case of environmental obstructions, moreover, it might be vital to numerically model the blockage penetration loss as a function of the material (e.g., standard multi-pane glass, concrete, wood, etc.) and the size of the blocker, as it usually occurs in the cellular context~\cite{38901}.

  \item \emph{Vehicular traffic density.} 
  %The distance dependence of the path loss should vary with the surrounding environment. 
  Currently, the 3GPP model does not distinguish between different densities of vehicular traffic.
  Also considering the extended model in~\cite{boban2016modeling}, Fig.~\ref{fig:Pathloss}c demonstrates that the difference between density regimes is limited (i.e., in the order of 2-3 dB) and, in case of urban propagation, only affects short-range communications.
  On the other hand,  uncrowded roads have been proven to have the lowest path loss and path loss variation among all the environments (e.g., the increase in the path loss variation  due to heavier traffic on crowded roads was experimentally measured to be above 10 dB~\cite{takahashi2003distance}). 
  Density-dependent path loss characterizations should therefore be provided.

  \item \emph{Blockage Characteristics and Number of Blockers.}  
  While in the 3GPP model the size of the obstructing vehicle has  negligible effects (see Fig.~\ref{fig:Pathloss}b), the blockage loss difference between small and large vehicles has been shown to range from 1 to 6 dB at high frequencies~\cite{boban2019multi}.
  %, depending on the location of the intermediate vehicle between the transmitter and the receiver. 
  Moreover, while the 3GPP considers a single intermediate vehicle located midway between the transmitting and the receiving vehicle in case of NLOSv, adding a second blocking vehicle in addition to the first one might lead to significantly increased blockage loss (e.g., more than 5 dB at 60 GHz~\cite{R11801398}). 
  % In case of environmental obstructions, it might also be vital to numerically model the blockage penetration loss as a function of the material (e.g., standard multi-pane glass, concrete, wood, etc.) and the size of the blocker, as it usually occurs in the cellular context~\cite{38901}.
 Such effects should be included in V2V channel models.

  \item \emph{Antenna Placement.} Although a rooftop position for the vehicle antennas was observed to provide optimal coverage, at mmWaves the self-blockage effect caused by the curvature of the roof raises questions about whether  rooftop positions are preferable  to  bumper locations~\cite{mecklenbrauker2011vehicular}.

  \item \emph{Temporal and Spatial Correlation.} The lack of temporally and spatially correlated channel measurements in the mmWave band significantly limits the level of detail that can be achieved in simulations, as it becomes impossible to make a clear assessment about how the dynamics of the channel affect the network performance.
  %Due to the lack of temporally and spatially correlated channel measurements in the mmWave band, it is currently not possible to develop an accurate statistical model for mobility-related scenarios. This significantly limits the level of realism and detail that can be achieved in simulations, as it becomes impossible to make a clear assessment about how the dynamics of the channel affect the network performance.
  % Moreover, the 3GPP channel model assumes independent drops of vehicles in space and time, therefore time evolution and spatial consistency of the LOS blockage probability are not captured.

  %Moreover, time- and space-consistent evolution of LOS blockage based on the location of the transmitter and the receiver and the composition of their surroundings has important implications to assign the appropriate path loss, shadowing, small-scale, and large-scale parameters over time and space

  \item \emph{Use of Directional Antennas.}
  % MmWave links are typically directional to benefit from beamforming gain and  require precise alignment of the transmitter and receiver beams to maintain connectivity.
  %  affect the power angular profile and, in turn, the Doppler spread. 
 The effects of directional transmissions have not been numerically characterized  by currently available channel measurements, which make use of isotropic antennas with the assumption of unit gain or of horn antennas with fixed pointing direction.
It has also been reported that the delay spread decreases with narrow beams, but measurements in a vehicular context are  lacking~\cite{va2017impact}.

\item \emph{Fading Statistics.}
The statistics of the small-scale fading at mmWave frequencies are vital to  describe the fluctuations of the received power over time and to model deviations from the power predicted by the simple path loss equations. 
Such measurements also enable the study of the correlation among signals in a multipath environment, which is known to have significant impact on the system performance.
Most existing measurement campaigns model large-scale path loss parameters, while a complete characterization of the fading  statistics in a vehicular environment has  received little attention so far and deserves further investigation.

  %Such an effect has not been characterized yet by currently available channel measurements.
\end{itemize}

\vspace{-0.33cm}
\bibliographystyle{IEEEtran}
\bibliography{bibliography.bib}

\begin{thebibliography}{10}
\providecommand{\url}[1]{#1}
\csname url@samestyle\endcsname
\providecommand{\newblock}{\relax}
\providecommand{\bibinfo}[2]{#2}
\providecommand{\BIBentrySTDinterwordspacing}{\spaceskip=0pt\relax}
\providecommand{\BIBentryALTinterwordstretchfactor}{4}
\providecommand{\BIBentryALTinterwordspacing}{\spaceskip=\fontdimen2\font plus
\BIBentryALTinterwordstretchfactor\fontdimen3\font minus
  \fontdimen4\font\relax}
\providecommand{\BIBforeignlanguage}[2]{{%
\expandafter\ifx\csname l@#1\endcsname\relax
\typeout{** WARNING: IEEEtran.bst: No hyphenation pattern has been}%
\typeout{** loaded for the language `#1'. Using the pattern for}%
\typeout{** the default language instead.}%
\else
\language=\csname l@#1\endcsname
\fi
#2}}
\providecommand{\BIBdecl}{\relax}
\BIBdecl

\bibitem{Boccardi}
F.~Boccardi, R.~W. Heath, A.~Lozano, T.~L. Marzetta, and P.~Popovski, ``Five
  disruptive technology directions for 5{G},'' \emph{IEEE Communications
  Magazine}, vol.~52, no.~2, pp. 74--80, February 2014.

\bibitem{Heath_surveyV2X}
V.~Va, T.~Shimizu, G.~Bansal, and R.~W. Heath, ``Millimeter wave vehicular
  communications: A survey,'' \emph{Foundations and Trends® in Networking},
  vol.~10, no.~1, pp. 1--113, 2016.

\bibitem{3GPP_RP181480}
{3GPP}, ``{New SID on NR V2X (Release 16)},'' \emph{RP-181480}, 2018.

\bibitem{3GPP_181435}
------, ``{Study on NR beyond 52.6 GHz (Release 16)},'' \emph{RP-181435}, 2018.

\bibitem{rappaport2014millimeter}
T.~S. Rappaport, R.~W. Heath~Jr, R.~C. Daniels, and J.~N. Murdock,
  \emph{Millimeter wave wireless communications}.\hskip 1em plus 0.5em minus
  0.4em\relax Pearson Education, 2014.

\bibitem{MOCAST_2017}
M.~Giordani, A.~Zanella, and M.~Zorzi, ``{Millimeter wave communication in
  vehicular networks: Challenges and opportunities},'' in \emph{6th
  International Conference on Modern Circuits and Systems Technologies
  (MOCAST)}, May 2017.

\bibitem{kato2001its}
A.~Kato, K.~Sato, and M.~Fujise, ``{ITS wireless transmission technology.
  Technologies of millimeter-wave inter-vehicle communications: Propagation
  characteristics},'' \emph{Journal of the Communications Research Laboratory},
  vol.~48, pp. 99--110, March 2001.

\bibitem{takahashi2003distance}
S.~Takahashi, A.~Kato, K.~Sato, and M.~Fujise, ``{Distance dependence of path
  loss for millimeter wave inter-vehicle communications},'' in \emph{IEEE 58th
  Vehicular Technology Conference (VTC-Fall)}, Oct 2003.

\bibitem{yamamoto2008path}
A.~Yamamoto, K.~Ogawa, T.~Horimatsu, A.~Kato, and M.~Fujise, ``{Path-loss
  prediction models for intervehicle communication at 60 GHz},'' \emph{IEEE
  Trans. on Vehicular Technology}, vol.~57, no.~1, pp. 65--78, Jan 2008.

\bibitem{38901}
3GPP, ``{Study on channel model for frequencies from 0.5 to 100 GHz (Release
  14)},'' TR 38.901, 2018.

\bibitem{molisch2009survey}
A.~F. Molisch, F.~Tufvesson, J.~Karedal, and C.~F. Mecklenbrauker, ``{A survey
  on vehicle-to-vehicle propagation channels},'' \emph{IEEE Wireless
  Communications}, vol.~16, no.~6, pp. 12--22, Dec 2009.

\bibitem{37885}
3GPP, ``{Study on evaluation methodology of new Vehicle-to-Everything V2X use
  cases for LTE and NR (Release 15)},'' TR 37.885, 2019.

\bibitem{boban2016modeling}
M.~Boban, X.~Gong, and W.~Xu, ``{Modeling the evolution of line-of-sight
  blockage for V2V channels},'' in \emph{IEEE 84th Vehicular Technology
  Conference (VTC-Fall)}, 2016.

\bibitem{SUMO2012}
D.~Krajzewicz, J.~Erdmann, M.~Behrisch, and L.~Bieker, ``Recent development and
  applications of {SUMO - Simulation of Urban MObility},'' \emph{International
  Journal on Advances in Systems and Measurements}, vol.~5, no. 3\&4, pp.
  128--138, December 2012.

\bibitem{36885}
3GPP, ``{Study on LTE-based V2X services (Release 14)},'' TR 36.885, 2016.

\bibitem{boban2019multi}
M.~{Boban}, D.~{Dupleich}, N.~{Iqbal}, J.~{Luo}, C.~{Schneider}, R.~{Müller},
  Z.~{Yu}, D.~{Steer}, T.~{Jämsä}, J.~{Li}, and R.~S. {Thomä}, ``{Multi-Band
  Vehicle-to-Vehicle Channel Characterization in the Presence of Vehicle
  Blockage},'' \emph{IEEE Access}, vol.~7, pp. 9724--9735, Jan 2019.

\bibitem{R11801398}
{3GPP}, ``{V2X sidelink measurement results},'' {Huawei, HiSilicon -- Tdoc
  R1-1801398}, 2018.

\bibitem{mecklenbrauker2011vehicular}
C.~F. Mecklenbrauker, A.~F. Molisch, J.~Karedal, F.~Tufvesson, A.~Paier,
  L.~Bernad{\'o}, T.~Zemen, O.~Klemp, and N.~Czink, ``{Vehicular channel
  characterization and its implications for wireless system design and
  performance},'' \emph{Proceedings of the IEEE}, vol.~99, no.~7, pp.
  1189--1212, Feb 2011.

\bibitem{va2017impact}
V.~Va, J.~Choi, and R.~W. Heath, ``{The impact of beamwidth on temporal channel
  variation in vehicular channels and its implications},'' \emph{IEEE
  Transactions on Vehicular Technology}, vol.~66, no.~6, pp. 5014--5029,
  November 2017.

\end{thebibliography}

\end{document}